\documentclass[10pt]{article}
\usepackage[utf8]{inputenc}
\usepackage[T1]{fontenc}
\usepackage{amsmath}
\usepackage{amsfonts}
\usepackage{amssymb}
\usepackage[version=4]{mhchem}
\usepackage{stmaryrd}
\usepackage{graphicx}
\usepackage[export]{adjustbox}
\graphicspath{ {./images/} }
\usepackage{hyperref}
\hypersetup{colorlinks=true, linkcolor=blue, filecolor=magenta, urlcolor=cyan,}
\title{Performance Analysis of Monte Carlo Algorithms in Dense Subgraph Identification }

\author{Wanru Guo ${ }^{1, *}$\\
${ }^{1}$ Department of Biostatistics and Data Science, University of Texas\\
Health Science Center at Houston\\
${ }^{*}$ Corresponding author: wanru.guo@uth.tmc.edu}
\date{}

\begin{document}
\maketitle

\begin{abstract}
The exploration of network structures through the lens of graph theory has become a cornerstone in understanding complex systems across diverse fields. Identifying densely connected subgraphs within larger networks is crucial for uncovering functional modules in biological systems, cohesive groups within social networks, and critical paths in technological infrastructures. The most representative approach, the SM algorithm, cannot locate subgraphs with large sizes, therefore cannot identify dense subgraphs; while the SA algorithm previously used by researchers combines simulated annealing and efficient moves for the Markov chain. However, the global optima cannot be guaranteed to be located by the simulated annealing methods including SA unless a logarithmic cooling schedule is used. To this end, our study introduces and evaluates the performance of the Simulated Annealing Algorithm (SAA), which combines simulated annealing with the stochastic approximation Monte Carlo algorithm. The performance of SAA against two other numerical algorithms-SM and SA, is examined in the context of identifying these critical subgraph structures using simulated graphs with embeded cliques. We have found that SAA outperforms both SA and SM by 1) the number of iterations to find the densest subgraph and 2) the percentage of time the algorithm is able to find a clique after 10,000 iterations, and 3) computation time. The promising result of the SAA algorithm could offer a robust tool for dissecting complex systems and potentially transforming our approach to solving problems in interdisciplinary fields.
\end{abstract}

\section*{1 Introduction}
Network analysis provides invaluable insights into the complex interconnections and dynamics within various systems, enabling a deeper understanding of their\\
structure, behavior, and vulnerabilities across disciplines ranging from sociology and biology to technology and economics.Identifying densely connected subgraphs within a larger network has significant implications across numerous domains. From unraveling the intricate web of social interactions to understanding complex biological systems, the applications of such analyses are profound and far-reaching. The identification process pinpoints clusters or 'communities' that exhibit a high degree of interconnectivity, offering insights into the underlying structure and function of the network (Newman, M. E. J., 2003, Fortunato, S. 2010). The identification of densely connected subgraphs is an important research topic with broad applications in different fields, including social science, biology, computer science, psychology and finance (Barabási, A.-L., et al, 2011, Hartwell, L. H., 1999, Wasserman and Faust 1994). In biomedical sciences in particular, this analysis is crucial for detecting functional modules within protein-protein interaction networks, which may be pivotal in understanding cellular processes and disease mechanisms (Everett et al. 2006, Spirin and Mirny 2003).

In network analysis, a densely connected subgraph refers to a group of vertices that are highly connected in terms of vertex/edge connectivity. Let us denote a graph of network as $G(V, E)$, where $V$ be a set of $n$ nodes and $E$ be a set of $m$ edges in the network. Then, the edge density of $G(V, E)$ is defined as

\begin{equation*}
D(G)=\frac{2 m}{n(n-1)} \tag{1}
\end{equation*}

Obviously, the edge density $D$ defined in (1) characterizes how densely a graph is connected and it can take values within the range from 0 to 1 . When $D=1$, a graph is fully connected and such a graph is usually referred to as a clique in network analysis. When $D=0$, we have an empty graph with all nodes disconnected. Thus, the edge density can be used to evaluate how densely a graph is connected. Our goal is to find the densest subgraph with a fixed size $k$ that has the highest density $D$, where $k$ refers to the number of nodes in a subgraph. Clearly, this is equivalent to solving the following optimization problem:

\begin{equation*}
\arg \max \{D(S): S \subseteq G \text { and } S \text { has } k \text { nodes }\} \tag{2}
\end{equation*}

where $S$ denotes a subgraph in $G(V, E)$ with size $k$. The optimization in (2) is quite challenging because the number of subgraphs with size $k$ in a graph with $n$ nodes is $\binom{n}{k}$, which could be very large even for a moderate $n$.

In the network analysis literature, many statistical and computational methods have been proposed to identify the subgraph that is closely connected. The most representative method, suggested by Spirin and Mirny (2003), is designed for finding highly connected clusters of proteins in a network of protein interactions. This method is denoted as SM hereinafter. To describe the SM algorithm, let us define $L(S)=\sum_{i=1}^{k} \sum_{j=1}^{k} L_{i j}$, for any subgraph $S=\left\{v_{1}, \ldots, v_{k}\right\}$ with $k$\\
nodes, where $L_{i j}$ is the shortest path between node $v_{i}$ and node $v_{j}$ in graph $G$. The brief description of the SM algorithm is then given as follows.

\section*{SM Algorithm for Identifying the Densest Subgraph}
(i) Start with a set of $k$ nodes $S_{0}$.

(ii) For $l \geq 1$, in the $l$-th iteration, implement the following three steps:

(a) Choose a node $u_{0}$ randomly from $S_{l-1}$ and select a node randomly from the neighbors of $S_{l-1} \backslash u_{0}$ in $G$.

(b) Define $S_{l-1}^{\prime}=S_{l-1} \cup u_{1} \backslash u_{0}$. Set $S_{l}=S_{l-1}^{\prime}$ with probability $p$ and $S_{l}=S_{l-1}$ with probability $1-p$, where

$$
p= \begin{cases}1, & \text { if } L\left(S_{l-1}^{\prime}\right) \leq L\left(S_{l-1}\right) \\ \exp \left[-\frac{L\left(S_{l-1}^{\prime}\right)-L\left(S_{l-1}\right)}{k}\right], & \text { otherwise. }\end{cases}
$$

(c) Every ninth step, replace a node in $S_{l}$ with a node that is not connected to $S_{l}$.

(iii) Stop after the algorithm reaches a predetermined number of iterations.

After the algorithm stops, the graph in $\left\{S_{l}\right\}$ with the largest $D$ value is the densely connected subgraph that is identified by the SM algorithm. The SM algorithm performs well when $k$ is small (e.g., $k \leq 7$ ). But it has trouble locating dense subgraphs (Zhang and Chen 2015).

To overcome this limitation, Zhang and Chen (2015) proposed the SA algorithm for identifying subgraphs with the largest edge density $D$ for any size $k$. This algorithm combines the the ideas of simulated annealing (cf., Bertsimas and Tsitsiklis 1993) and efficient moves for the Markov chain. This algorithm, denoted as SA, can converge to the densest subgraph with probability one. In particular, they designed a proposal distribution that could increase the efficiency of the simulated annealing algorithm. A simulated annealing algorithm usually starts with a high temperature which ensures freedom in the transition between different states. Then, the temperature gradually decreases to a very small value based on a specific cooling schedule, which constrains the Markov chain moves within a small range of the objective value. At a fixed temperature $T$ in the simulated annealing algorithm, their proposal distribution for the Markov chain consists of two types of moves: the local move and the global move. Suppose the current state of the Markov chain is $S_{l-1}$. In the local move, we first randomly choose a node $u_{1}$ from the neighbors of $S_{l-1}$ and randomly choose a node $u_{0}$ from $S_{l-1}$

whose removal will not disconnect $S_{l-1} \cup u_{1}$. Denote $S_{l-1}^{\prime}=S_{l-1} \cup u_{1} \backslash u_{0}$ as the proposed state. In the global move, we first randomly choose a node $v_{1}$ which is not included in $S_{l-1}$. Then, for $i=2, \ldots, k$, we randomly select a node $v_{i}$ from the neighbors of $\left\{v_{1}, \ldots, v_{i-1}\right\}$. Denote $S_{l-1}^{\prime}=\left\{v_{1}, \ldots, v_{k}\right\}$ as the proposed state.

The following is the details of the SA algorithm:

\section*{SA Algorithm for Identifying the Densest Subgraph}
(i) Start with a set of $k$ nodes $S_{0}$.

(ii) For $l \geq 1$, in the $l$-th iteration, implement the following two steps:

(a) With probability of $\alpha$, propose a local move; with probability of $1-\alpha$, propose a global move. Denote the proposed state as $S_{l-1}^{\prime}$.

(b) Set $S_{l}=S_{l-1}^{\prime}$ with probability of $p$ and set $S_{l}=S_{l-1}$ with probability $1-p$, where $p=\min \left\{1, \exp \left[\frac{D\left(S_{l-1}^{\prime}\right)-D\left(S_{l-1}\right)}{T_{l}}\right]\right\}$, where $T_{l}$ is the temperature parameter at the $l$-th iteration.

(iii) Stop after the algorithm reaches a predetermined number of iterations.

In the above algorithm, $\alpha$ is the probability of local move in each iteration and $\left\{T_{l}\right\}$ is called the cooling schedule. Based on extensive simulation studies, Zhang and Chen (2015) suggested choosing $\alpha=0.9$ and we adopt this suggestion in this report. Regarding the cooling schedule, we choose to use the recommended geometric cooling system $T_{l}=0.001^{l / 1000}$ by Zhang and Chen $(2015)$.

As known by many researchers, the global optima cannot be guaranteed to be located by the simulated annealing methods including SA unless a logarithmic cooling schedule is used. However, the logarithmic cooling schedule is so slow that no one can afford to use this much CPU time. SA requires the logarithm cooling schedule to guarantee that the densest subgraph can be identified and thus the computation is extensive. This question came to mind, can we modify the SA algorithm to make it more efficient? In this report, we use a modified version of SA algorithm, known as SAA, and evaluate the efficiency of SAA with that of SM and SA algorithms through a simulation study.

\section*{2 Materials and Methods}
In the attempt to improve the efficiency of identifying densely connected subgraphs, we used the idea of simulated stochastic approximation annealing (Liang et al. 2014). The modified algorithm is denoted as SAA in this report. In the results section, we illustrate the performance of the SA algorithm using a simulated example and we compare the performance of the SM, SA and SAA in a simulation study. The SAA algorithm is a combination of simulated annealing and the stochastic approximation Monte Carlo algorithm. Under the framework of stochastic approximation, it is shown that SAA can work well with a cooling schedule in which the temperature can decrease much faster

than that in the logarithmic cooling schedule, for example, a square-root cooling schedule, while guaranteeing the global optima to be reached when the temperature tends to zero.

The algorithm is as follows. Let $E_{1}, \ldots, E_{N}$ be a partition of the sample space based on the value of $D$ :

$$
\begin{aligned}
E_{1} & =\left\{S: D(S) \leq a_{1}\right\}, E_{2}=\left\{S: a_{1}<D(S) \leq a_{2}\right\}, \ldots \\
E_{N-1} & =\left\{S: a_{N-2}<D(S) \leq a_{N-1}\right\}, E_{N}=\left\{S: D(S)>a_{N-1}\right\}
\end{aligned}
$$

where $a_{1}<\cdots<a_{N-1}$ are pre-specified numbers. The SAA algorithm is described below.

\section*{SAA Algorithm for Identifying the Densest Subgraph}
(i) Start with a set of $k$ nodes $S_{0}$.

(ii) For $l \geq 1$, in the $l$-th iteration, implement the following three steps:

(a) With probability of $\alpha$, propose a local move; with probability of $1-\alpha$, propose a global move. Denote the proposed state as $S_{l-1}^{\prime}$.

(b) Set $S_{l}=S_{l-1}^{\prime}$ with probability of $p$ and set $S_{l}=S_{l-1}$ with probability $1-p$, where $p=\min \left\{1, \exp \left[\frac{D\left(S_{l-1}^{\prime}\right)-D\left(S_{l-1}\right)}{T_{l}}+\theta_{J\left(S_{l-1}\right)}^{l-1}-\theta_{J\left(S_{l-1}^{\prime}\right)}^{l-1}\right]\right\}$, where $T_{l}$ is the temperature parameter at the $l$-th iteration and $J(\cdot)$ is a function such that $J(x)=i$ if $x \in E_{i}$.

(c) Set $\boldsymbol{\theta}^{l}=\boldsymbol{\theta}^{l-1}+\eta_{l}\left(\boldsymbol{e}_{l}-\boldsymbol{\pi}\right)$, where $\boldsymbol{\theta}^{l}=\left(\theta_{1}^{l}, \ldots, \theta_{N}^{l}\right)^{T}, \eta_{l}$ is called the gain factor at the $l$-th iteration, $\boldsymbol{e}_{l}=\left(I\left(S_{l} \in E_{1}\right), \ldots, I\left(S_{l} \in E_{N}\right)\right)^{T}, I(\cdot)$ is the indicator function and $\boldsymbol{\pi}$ is a pre-specified desired sampling distribution.

(iii) Stop after the algorithm reaches a predetermined number of iterations.

To use SAA, the parameters $N$ and $\left\{a_{i}: i=1, \ldots, N-1\right\}$, the cooling schedule $\left\{T_{l}\right\}$, the gain factor sequence $\left\{\eta_{l}\right\}$, the desired sampling distribution $\boldsymbol{\pi}=\left(\pi_{1}, \ldots, \pi_{N}\right)^{T}$ and the probability of local move $\alpha$ need to be chosen properly. By the suggestion in Liang et al. (2014), we have the following recommendations for their choices:

\begin{itemize}
  \item The sample space can be divided into $N=51$ subregions.
  \item The parameters $\left\{a_{i}\right\}$ can be chosen by trial and error. In simulations, we usually choose $a_{1}$ so small that $E_{1}$ is an empty set, and choose $a_{N-1}$ so large that SAA can quickly move out from $E_{N}$ to other subregions.
  \item The square root cooling scheme: $T_{l}=0.001 \sqrt{\frac{1500}{\max (l, 1500)}}$.
  \item The gain factor sequence: $\eta_{l}=1500 / \max (1500, l)$.
  \item The desired sampling distribution: $\pi_{i}=\frac{\exp ^{-0.1(i-1)}}{\sum_{j=1}^{N} \exp ^{-0.1(j-1)}}$, for $i=1, \ldots, N$.
  \item The probability of local move is set to be $\alpha=0.9$.
\end{itemize}

\section*{3 Results}
In this part, we investigate the performance of SM, SA and SAA using a Monte Carlo simulation study. Before presenting the simulation results, let us provide a detailed description about the simulation setup. We first generated a random graph with $n=100$ nodes and each edge is generated with a probability of 0.05 . Then, we embedded a size 10 clique in the graph. Figure 1 shows the graph we generated. From the figure, we can see that it is very hard to visually identify the embedded clique in the graph.

\begin{center}
\includegraphics[max width=\textwidth]{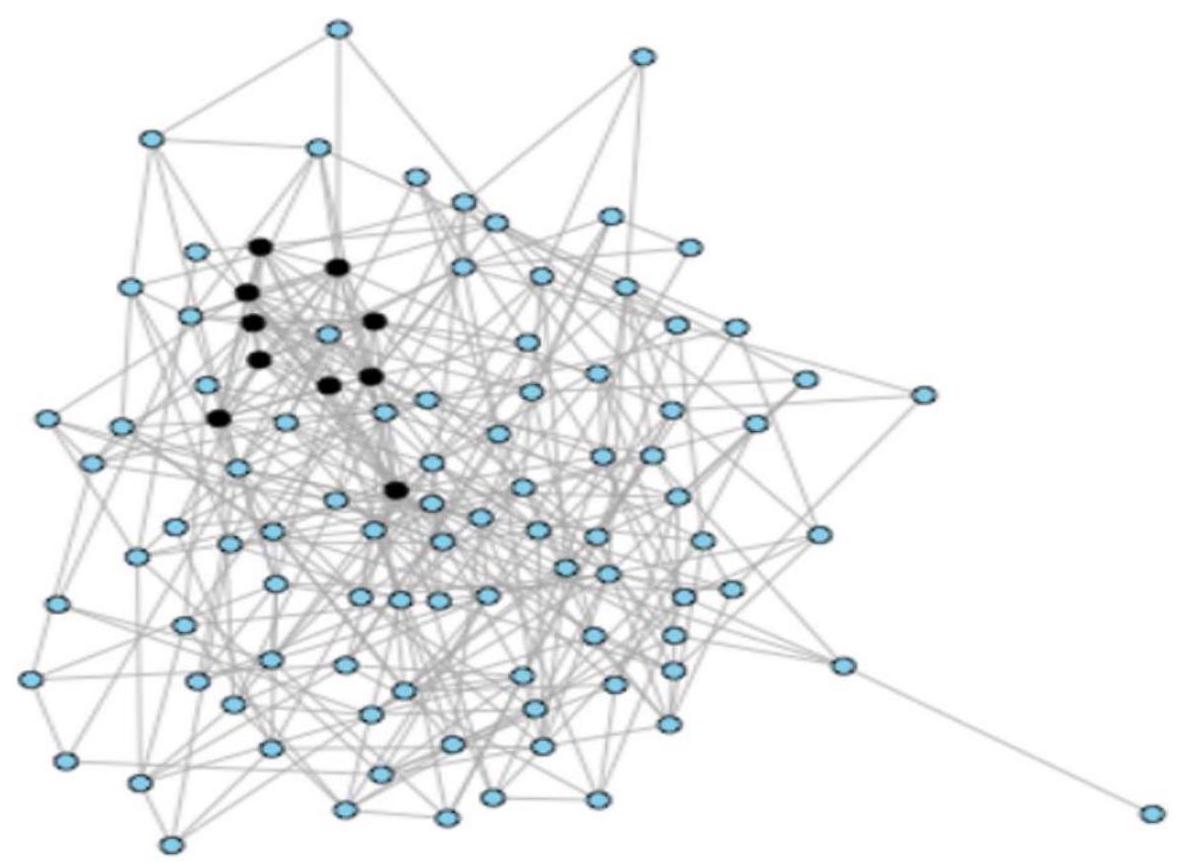}
\end{center}

Figure 1: The simulated graph with an embedded size 10 clique. The nodes in the embedded clique are in black.

We applied the three algorithms to search for the densest subgraph of size 10 in Figure 1. The trace plots of the edge density for the three algorithms are presented in Figures 2-4. As can be seen from Figure 2, SM cannot identify the size 10 clique after 10,000 iterations. This is consistent with our expectation. Because SM cannot work well when $k$ is greater than 7 . From Figures $3-4$, it is clear that (i) both SA and SAA can find the clique after 10,000 iterations, and (ii) $\mathrm{SA}$ requires more than 8,000 iterations to identify the densest subgraph while SAA only needs about 3,000 iterations. So, in this example SAA is preferable in terms of the number of iterations required to find the densest subgraph.

\begin{center}
\includegraphics[max width=\textwidth]{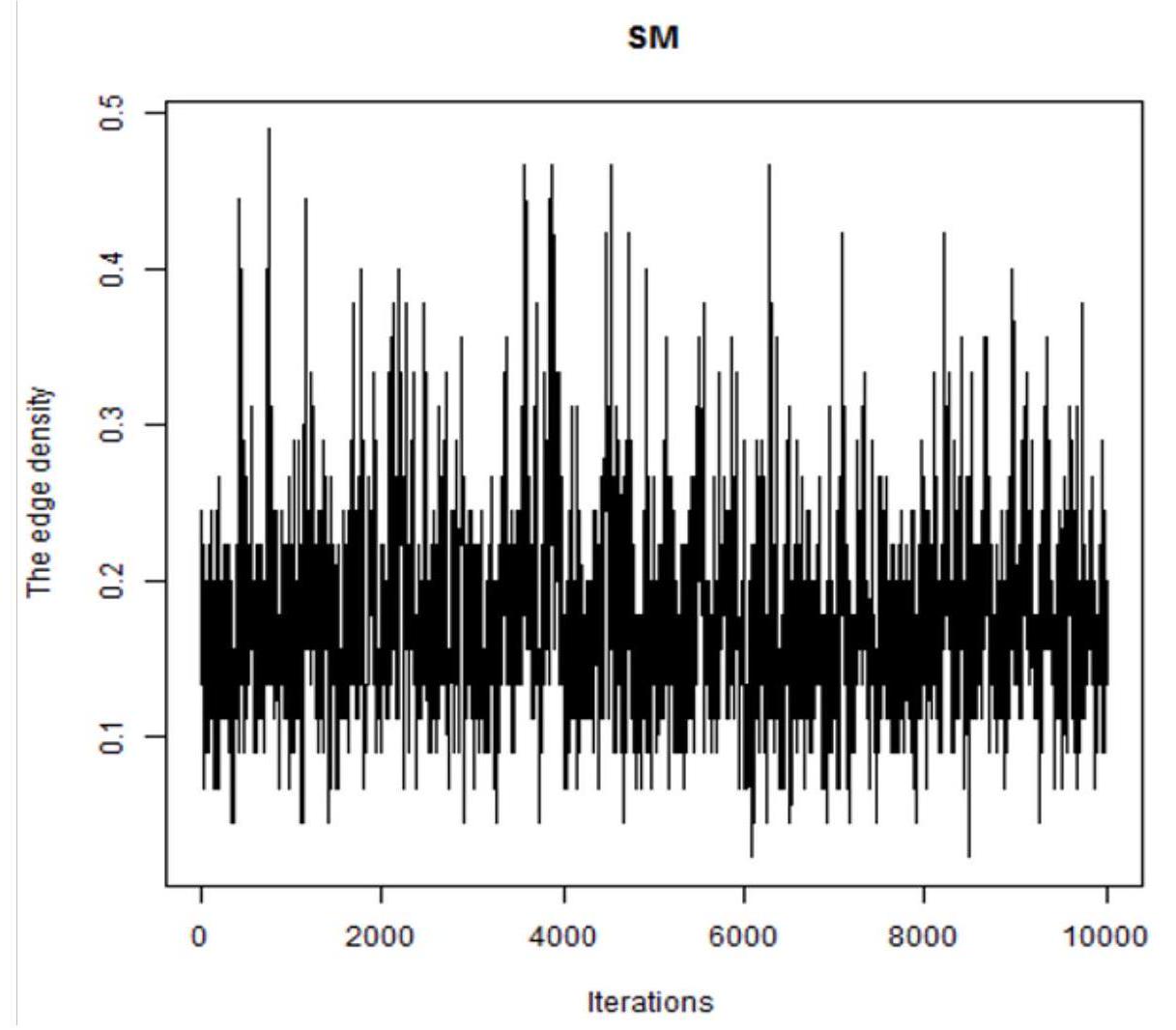}
\end{center}

Figure 2: The trace plot of the edge density $D$ for the SM algorithm.

Finally, we repeated the above procedure 100 times by generating 100 graphs and embedding a size 10 clique in each of the graphs. SM can only find the clique 12 out of the 100 times after 10,000 iterations. SA identified the densest subgraph in 84 cases and SAA found the clique in all 100 cases. Regarding the computation time for identifying the

\begin{center}
\includegraphics[max width=\textwidth]{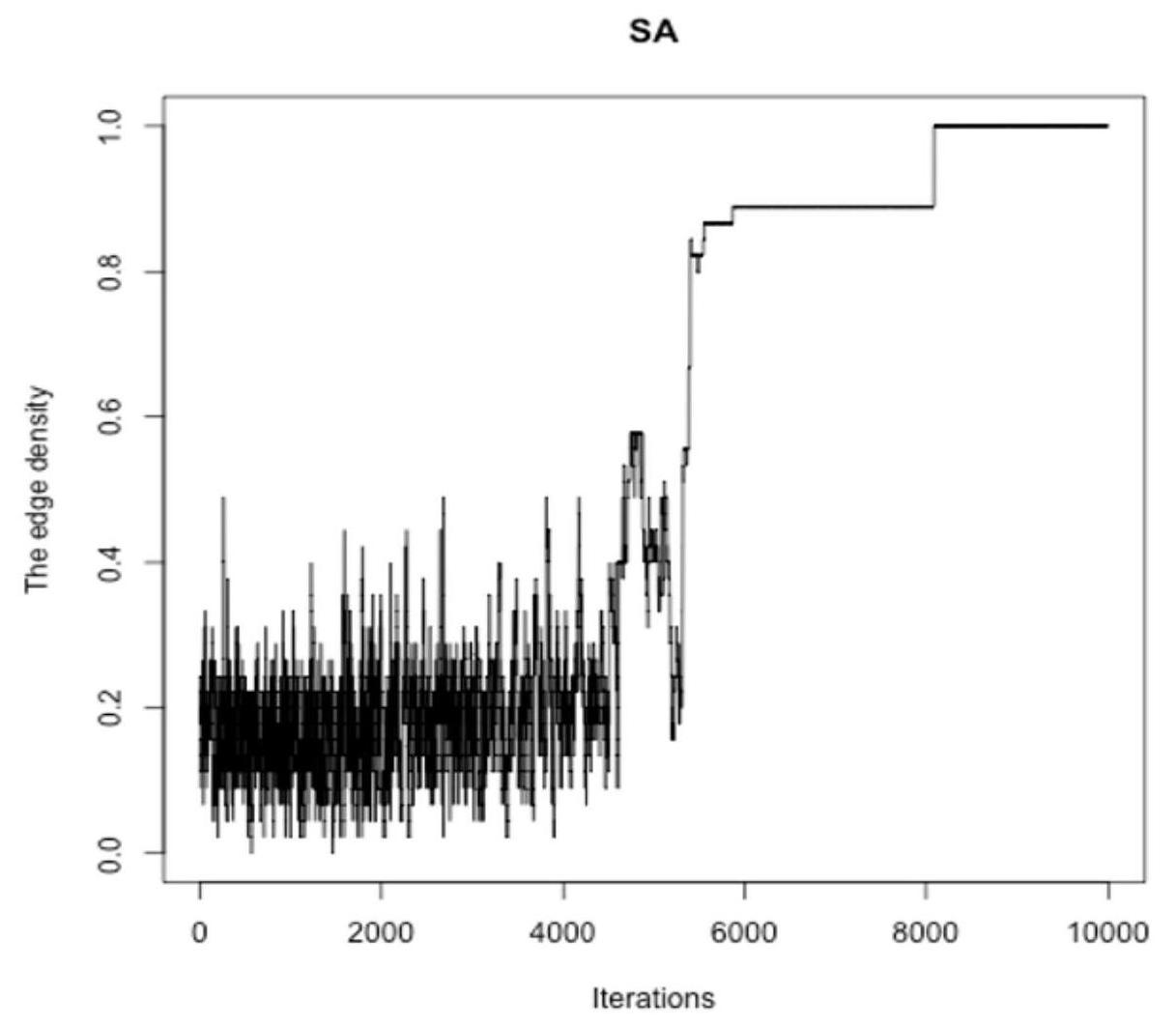}
\end{center}

Figure 3: The trace plot of the edge density $D$ for the SA algorithm.

clique, the average CPU times for SM, SA and SAA are 583 seconds, 312 seconds and 119 seconds, respectively, when running code on a Mac desktop with a $2.9 \mathrm{GHz}$ Intel Core i5 processor. So, this example confirms the benefit of using SAA.

\begin{center}
\includegraphics[max width=\textwidth]{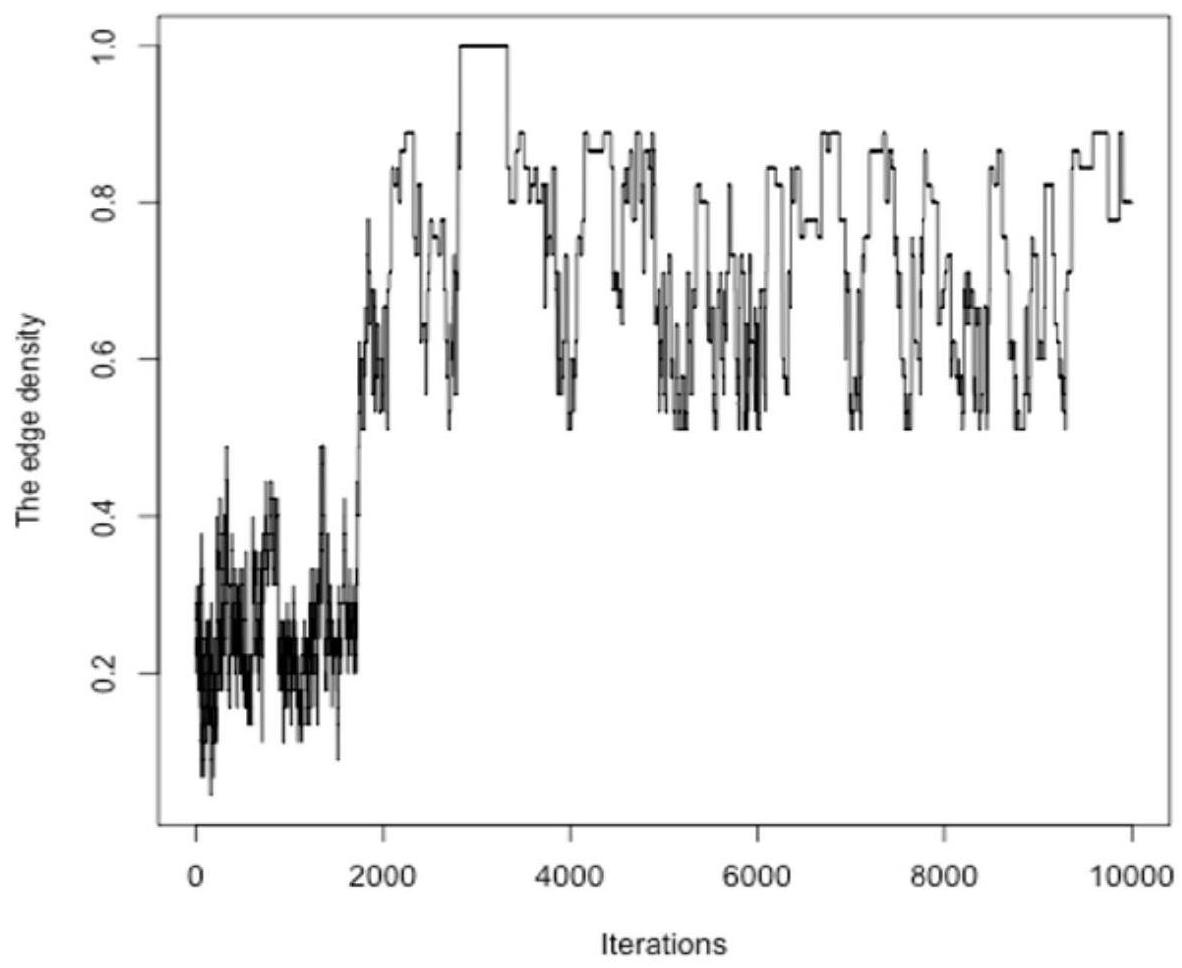}
\end{center}

Figure 4: The trace plot of the edge density $D$ for the SAA algorithm.

\section*{4 Discussion}
The SA algorithm was proposed by combining the ideas of simulated annealing and efficient moves for the Markov chain. Under some regularity conditions, the SA algorithm was shown to converge to the densest subgraph with probability one. Besides, extensive simulation studies and the application to two real data examples including a yeast protein interaction network and a stock market graph demonstrates that SA can outperform the commonly used SM algorithm in many different cases (Zhang and Chen, 2015). Thus, the SA algorithm should provide a reliable tool for finding the most densely connected subgraph in a network. However, the global optima can only be guaranteed to be located by SA with the logarithm cooling schedule which is very slow and infeasible in many applications. To overcome this limitation, we can consider the SAA algorithm that can be used to speed up the convergence to the global optima. We have demonstrated the superiority of SAA compared to SA and SM in firstly, the efficiency in identifying cliques within a randomly generated graph in terms of the smallest number of iterations taken; secondly, having the highest percentage\\
of clique identification in a large number of simulations after 10,000 iterations; and lastly, shortest computation time.

The application of SAA is particularly promising in computational biology, especially in the analysis of protein-protein interaction networks. These networks are crucial in understanding the complex interplay of cellular mechanisms. By applying the SAA algorithm, researchers can pinpoint highly interconnected protein clusters that often indicate functional complexes pivotal for various cellular processes (Spirin \& Mirny, 2003). Such identification is not only key to unraveling basic biological functions but also has implications in disease pathology and the development of targeted treatments (Aittokallio \& Schwikowski, 2006). Additionally, in structural biology, SAA's capabilities can assist in identifying stable motifs within protein structures, which are indicative of functional sites (Hartwell et al., 1999). This is particularly useful in drug discovery, where such motifs represent potential binding sites for therapeutic agents. The precision and efficiency of SAA in mining vast network data to uncover these vital biological interactions exemplify its value in propelling forward our understanding of protein functions and interactions, which is a step toward the acceleration of therapeutic innovation.

Beyond its promising applications in computational biology, the SAA algorithm exhibits significant potential in the realm of social network analysis-a field where understanding the dynamics and structure of connections can yield critical insights. In social networks, dense subgraphs often correspond to communities or clusters that are tightly knit and highly interactive, which are of particular interest for sociologists and anthropologists studying social cohesion and group dynamics (Wasserman \& Faust, 1994). The SAA algorithm, with its refined accuracy, could revolutionize the way we understand social structures, allowing for the identification of influential groups, information dissemination patterns, and the emergence of trends within both online and offline social communities. Furthermore, its application could extend to cybersecurity, where identifying dense network traffic subgraphs could assist in detecting unusual patterns associated with cyber threats or network vulnerabilities (Newman, 2003). In a world increasingly dependent on digital communication, the role of robust network analysis tools such as SAA becomes indispensable in safeguarding information integrity and understanding the fabric of digital interactions.

\section*{5 Conclusion}
In this paper, we introduce three numerical algorithms for identifying dense subgraphs including SM, SA and SAA. The three algorithms SM, SA and SAA are compared through a simulation study in this report, and the simulation results indicate that SAA outperforms both SA and SM in efficiently identifying densely connected subgraphs within our simulated network. The superiority of SAA algorithm suggests its potential for significant applications in areas that require the analysis of complex networks, with significant applications in the field of computational biology, particularly in the analysis of protein-protein\\
interaction networks.

\section*{References}
\begin{enumerate}
  \item Aittokallio, T., \& Schwikowski, B. (2006). Graph-based methods for analysing networks in cell biology. Briefings in Bioinformatics, 7(3), 243-255. \href{https://doi.org/10.1093/bib/bbl022}{https://doi.org/10.1093/bib/bbl022}

  \item Barabási, A.-L., Gulbahce, N., \& Loscalzo, J. (2011). "Network medicine: a network-based approach to human disease." Nature Reviews Genetics, 12(1), 56-68. doi:10.1038/nrg2918.

  \item Bertsimas, D., and Tsitsiklis, J. (1993), "Simulated Annealing,", Statistical Science, 8, 10-15.

  \item Everett, L., Wang, L., and Hannenhalli, S. (2006), "Dense Subgraph Computation via Stochastic Search: Application to Detect Transcriptional Modules," Bioinformatics, 22, 117-123.

  \item Fortunato, S. (2010). "Community detection in graphs." Physics Reports, $486(3-5), 75-174$. doi:10.1016/j.physrep.2009.11.002.

  \item Hartwell, L. H., Hopfield, J. J., Leibler, S., \& Murray, A. W. (1999). "From molecular to modular cell biology." Nature, 402 (6761 Suppl), C47-C52. doi: $10.1038 / 35011540$.

  \item Liang, F., Cheng, Y., and Lin, G. (2014), "Simulated Stochastic Approximation Annealing for Global Optimization With a Square-Root Cooling Schedule," Journal of the American Statistical Association, 109, 847-863.

  \item Newman, M. E. J. (2003). "The Structure and Function of Complex Networks." SIAM Review, 45(2), 167-256. doi:10.1137/S003614450342480.

  \item Spirin, V., and Mirny, L. A. (2003), "Protein Complexes and Functional Modules in Molecular Networks," Proceedings of the National Academy of Sciences, 100, 12123-12128.

  \item Wasserman, S. S., and Faust, A. K. (1994), Social Network Analysis: Methods and Applications, Cambridge: Cambridge University Press.

  \item Zhang, J., and Chen, Y. (2015), "Monte Carlo Algorithms for Identifying Densely Connected Subgraphs," Journal of Computational and Graphical Statistics, 24, 827-845.

\end{enumerate}

\end{document}